%% file: Final.tex
\documentclass[11pt,journal]{IEEEtran}
\IEEEoverridecommandlockouts

\usepackage{mathtools}
\usepackage{subfig}
\usepackage{nicefrac} 
\usepackage{graphicx}
\usepackage{array}
\usepackage{algorithm}
\usepackage{algorithmic}

\usepackage{cite}
\usepackage{amsmath, amssymb, amsthm,amsfonts} 
\usepackage[T1]{fontenc}               

\theoremstyle{definition}

\newcommand{\mc}[1]{\mathcal{#1}}

\begin{document}

\title{Energy-Efficient Secrecy in Wireless Networks Based on Random Jamming}
\author{
	Azadeh Sheikholeslami, Majid Ghaderi,\IEEEmembership{ Member, IEEE}, Hossein Pishro-Nik,\IEEEmembership{ Member, IEEE},  \\Dennis Goeckel,\IEEEmembership{ Fellow, IEEE}
	\thanks{This work was supported by the National Foundation under Grant CIF-1421957.}
}
\maketitle
\begin{abstract}

	\input{abstract.tex}
\end{abstract}


\section{Introduction}

\input{introduction.tex}
\section{System Model and Approach}\label{sec:system_and_approach}

\subsection{System Model}\label{sec:system}
We consider a wireless network with  nodes that are distributed arbitrarily.
A source node generates the message and conveys it to a destination node in a multi-hop fashion.
An $H$-hop path from the source to the destination  is denoted by $\Pi=\left\langle \ell_1,\ldots,\ell_H\right\rangle$, where $\ell_i$ is the link that connects two nodes $S_i$ and $D_i$ along the path $\Pi$.
There are also  non-colluding eavesdroppers present in the network such that the message transmission of each link is prone to be overheard by multiple eavesdroppers. We denote the set of eavesdroppers by $\mc{E}$.
The eavesdroppers are assumed to be passive, and thus their locations and their channel-state informations  are not known to the legitimate nodes.
We assume that the system nodes are equipped with omni-directional antennas while the eavesdroppers can  be equipped with more sophisticated directional antennas.

For the channel, we consider transmission in a quasi-static Rayleigh fading environment. 
Let $h_{S,D}$ be the fading coefficient between node $S$ and node $D$ (This assumption is relaxed for eavesdroppers' channels, as discussed later.).
Without loss of generality, we assume $E[|h_{S,D}|^2]=1$.
Suppose the transmitter $S$ transmits the signal $x_S$ at power level $P_S$.
The signal that the receiver $D$ (analogously,  eavesdropper $E$) receives is:
\[\tilde{y}_D=\frac{x_S h_{S,D}}{d_{S,D}^{\frac{\alpha}{2}}}+n_D
\]
where $d_{S,D}$ is the distance between $S$ and $D$, $\alpha$ is the path-loss exponent, and $n_D\sim\mathcal{N}\left(0,\sigma_D^2\right)$ is additive white Gaussian noise (AWGN) at the receiver $D$.

Because compression of a receiver's front-end dynamic range is the biggest challenge when operating in the presence of strong jamming, we also consider the effect of the analog-to-digital converter (A/D) on the received signal, which consists of the quantization noise and the quantizer's overflow.
The quantization noise is a result of  the limited resolution of the A/D, and the quantizer's overflow happens when the range of the received signal is larger than the span of the A/D.
We assume that the quantization noise is uniformly distributed \cite[Section 5]{widrow2008quantization}. The resolution of a $b$-bit A/D with full dynamic range $[-r,r]$
is
\[\delta=\frac{2r}{2^b}.\]
Suppose the receiver  has a $b_D$-bit A/D and the eavesdropper has a $b_E$-bit A/D.
Since the power of the received signal at the receiver $D$ is $\frac{P_S|h_{S,D}|^2}{d_{S,D}^{\alpha}}$, we set the range of the A/D as,
\[
r_D=l\frac{\sqrt{P_S}|h_{S,D}|}{d_{S,D}^{\alpha/2}},
\]
 where $l$ is a constant that maximizes the mutual information between the transmitted signal and the received signal \cite{sheikholeslami2013everlasting}.
The resolution of the A/D of the receiver $D$ is:
\[
\delta_D=\frac{2l\sqrt{P_S}|h_{S,D}|}{2^{b_D}d_{S,D}^{\alpha/2}}.
\]
Analogously,  the range of the eavesdropper's A/D is,
\[
r_E=l\frac{\sqrt{P_S}|h_{S,E}|}{d_{S,E}^{\alpha/2}},
\]
and hence, the resolution of the A/D of the eavesdropper $E$  is:
\[
\delta_E=\frac{2l\sqrt{P_S}|h_{S,E}|}{2^{b_E}d_{S,E}^{\alpha/2}}.
\]
\subsection{Approach: Random Jamming for Secrecy}\label{sec:approach}

Our goal is to obtain end-to-end everlasting secrecy, which means that even if each eavesdropper
works forever on the signal that is recorded, it will not be able to extract the message.
Unlike cryptography, we do not assume any limitation on the computational capability of the
eavesdropper. Instead, we exploit \textit{current} hardware limitations of the eavesdropper to achieve
everlasting security, as  explained in detail in \cite{sheikholeslami2013everlasting,sheikholeslami2015jamming}.
At each hop, we use the random jamming scheme of \cite{sheikholeslami2014everlasting,sheikholeslami2015jamming} to provide everlasting secrecy.
In this scheme, based on a cryptographic key that is shared between the legitimate nodes, a jamming signal with large variation is added to the transmitted signal. 
It is assumed that the cryptographic key should be kept secret just for the time of transmission, and can be revealed to the eavesdropper  right after transmission without compromising secrecy.
The legitimate receiver can use its key to cancel the effect of the jamming before analog-to-digital-conversion (A/D), while the eavesdropper must record the signal and jamming, and cancel the effect of jamming  later from the recorded signal (after analog-to-digital-conversion).
Hence, the signal that the legitimate receiver receives is well-matched to its A/D converter. 
On the other hand, the large variation of the random jamming signal causes overflow of the eavesdropper's A/D.
The eavesdropper may enlarge the span of her A/D to prevent overflows; however, it degrades the resolution of its A/D, thus increasing the A/D noise.

Note that unlike cryptography, the secret key used in the random jamming approach \textit{only needs to be
	kept secret for the duration of the wireless transmission (i.e.
	it can be given to Eve immediately afterward)}.
The eavesdropper must store the signal and
try to cancel the jamming signal from the recorded signal at
the output of her A/D after she obtains the key.
However, the jamming signal is designed such that Eve has already lost
the information she would need to recover the secret message,
even if she obtains the key immediately after the transmission.
In order to gain more insight into the difference between this approach and cryptography, 
suppose that the legitimate receivers have access to a standard key exchange protocol that is currently computationally
secure in the near-term beyond any reasonable doubt (e.g. 1028-bit Elliptic Curve Diffie-
Hellman).
If we employ  the proposed scheme or cryptography to convey a secret message, we encounter two  risks, respectively:

\begin{enumerate}
	\item 
	Risk 1: In practice, 
	the time it takes to transmit a message over the wireless channel is 
	very short and in the order of a few milliseconds, e.g., 10 milliseconds.
	An eavesdropper records the key establishment messages and breaks that key
	in the next 10 milliseconds (the time during which we are using that key to transmit
	the message we desire to keep secret forever with our technique). Obviously, there
	would not be much technological advance in those 10 milliseconds, so he/she is
	essentially limited to 10 milliseconds of effort with the technology in place at the
	time of message transmission.
	\item
	 Risk 2: An eavesdropper records the key establishment messages and ciphertext of
	a standard cryptographic approach, and then uses an unlimited amount of time (say,
	20 years as a lower bound to unlimited) to break that key and decode the secret
	message. Obviously, the eavesdropper then  not only has a much longer time, but also can take advantage of what are certain to be significant technological
	advances in algorithms, computation, and methods of "hacking" the key from one of
	the parties.
\end{enumerate}
Clearly, Risk 1 and Risk 2 are very different risk classes, and one would feel much more comfortable with Risk 1 (which is presented by our scheme) than Risk 2 (which is the risk of standard cryptographic approaches) when attempting to achieve everlasting security.

In  \cite{sheikholeslami2014everlasting,sheikholeslami2015jamming} it is shown that, although increasing the span of the A/D causes the eavesdropper to suffer from more quantization noise, the overflows are more harmful, and thus the best strategy that the eavesdropper can employ is to enlarge the span of its A/D such that it captures all of the signal and thus no overflow occurs.

The random jamming signal  $J$ that the 	transmitter adds to its signal follows a uniform distribution with $2^K$ jamming levels. 
Hence, $K$ bits of the cryptographic key to generate each jamming symbol are needed. 
The distance between two consecutive jamming levels is $2l\sqrt{P_S}$.
Thus, the average energy that is spent on the random jamming  signal is,
\begin{align}
	\nonumber P_J&=E[J^2]\\
	\nonumber &=\frac{1}{2^K}\sum_{j=0}^{2^K-1}\left(2l\sqrt{P_S}j\right)^2\\
	\nonumber &=\frac{4l^2{P_S}}{2^K}\sum_{j=0}^{2^K-1}j^2\\
	\nonumber &=\frac{4l^2{P_S}}{2^K}\times\frac{2^{3K+1}-3\times2^{2K}+2^K}{6}\\
	\nonumber  &=\frac{2l^2\left(2^{2K+1}-3\times2^{K}+1\right)}{3}P_S\\
	 &=\beta P_S
	\label{eq:PJ} 
\end{align}
where $\beta$ is a constant that  depends on $K$. 

 Suppose that the eavesdropper uses a $b_E$-bit A/D.
Since the power of the signal at the eavesdropper's receiver is $\frac{P_S|h_{S,E}|^2}{d_{S,E}^{\alpha}}$, and considering the automatic-gain-control of the eavesdropper's receiver, the resolution of the eavesdropper's A/D before jamming is:
\begin{align}\label{eq:delta}
\delta_E=\frac{2l\sqrt{P_S|h_{S,E}|^2}}{2^{b_E}d_{S,E}^{\alpha/2}}
\end{align}
Now suppose that the transmitter adds the jamming to its signal. 
Since the eavesdropper does not know the key, it should enlarge the  span of its A/D to capture all the signal plus jamming. 
The maximum amplitude of the signal plus jamming can be written as,
\[\frac{\sqrt{P_S|h_{S,E}|^2}}{d_{S,E}^{\alpha/2}}+(2^K-1)\frac{\sqrt{P_S|h_{S,E}|^2}}{d_{S,E}^{\alpha/2}}=2^K\frac{\sqrt{P_S|h_{S,E}|^2}}{d_{S,E}^{\alpha/2}}\]
Thus, the resolution of eavesdropper's A/D is:
\begin{align}\label{eq:delta'}
	\delta'_E=\frac{2l\sqrt{P_S|h_{S,E}|^2}}{2^{b_E}d_{S,E}^{\alpha/2}}\times2^K=\frac{2l\sqrt{P_S|h_{S,E}|^2}}{2^{b_E-K}d_{S,E}^{\alpha/2}}
\end{align}

The random jamming scheme of \cite{sheikholeslami2014everlasting,sheikholeslami2015jamming} relies on the limited resolution of the eavesdropper's A/D.
As opposed to cryptography, technology improvement in the future are not of concern here because the signal cannot be captured. 
Hence, we should assume that the legitimate nodes  either know a bound  on the quality of the eavesdroppers' A/Ds, or plan for the case that all eavesdroppers use the best A/D technology available at the time.
The  realization of this assumption is facilitated by the fact that A/D technology progresses very slowly\footnote{For a complete discussion on this  see   \cite[Section V]{sheikholeslami2015jamming}.}.  
Hence, throughout this paper we assume that the resolution of the A/D of each  eavesdropper is equal to or less than $b_E$ bits.

\subsection{Jamming Cancellation at the Legitimate Receiver}

Nearly all techniques that exploit 
jamming for secrecy ignore the effects of channel estimation error (e.g. \cite{negi2005secret,tekin2008general,dong2009cooperative,krikidis2010jamming,zheng2011optimal,xie2014secure,gabry2015energy,ghaderi2014min,sheikholeslami2014everlasting,sheikholeslami2015jamming}), yet 
it is important since in real systems  the jamming power is high, and thus the residual jamming due to imperfections in channel estimation can be considerable.
Note that from \cite{hassibi2003much,semmelrodt2003investigation}, the channel estimation error might be 
very small, but, since we have high-power jammers, the residual 
interference is still important and can have an impact on system 
performance.
Hence,  we consider the residual jamming at the receiver due to errors 
in the channel estimates.  
Given a pilot-based approach for channel 
estimation, the channel estimate is conditionally 
Gaussian, where the mean of this Gaussian distribution is the minimum 
mean-squared estimate (MMSE) channel estimate. 
The estimation error of this MMSE   is a zero-mean Gaussian random variable   with variance $\theta^2$ which is a constant (e.g. see \cite{goeckel1999adaptive}).
The resultant noise is a multiplication of two independent Gaussian random variables; the residual channel estimation error and the received jamming signal. 
Hence, the channel estimation  noise is a zero-mean non-Gaussian random variable with variance,
\begin{equation}
\sigma_J^2=\theta^2\frac{P_J|h_{S,D}|^2}{d_{S,D}^{\alpha}}.
\end{equation}

\subsection{Metric}

Since the quantization noise is  uniformly distributed \cite[Section 5]{widrow2008quantization} and the channel estimation  noise is  non-Gaussian, the derivation of the capacity of the channel between transmitter and receiver, and the channel between transmitter and eavesdropper, is not straightforward.
Thus, we apply an upper-bound and a lower-bound of the capacity of a channel with independent additive noise as described in  \cite{shannon1949communication} and \cite{iha1978cap}.
Suppose that the resolution of the A/D of receiver $D$ is $\delta_D$.
The capacity of the channel between the transmitter $S$ and the receiver $D$ conditioned on the fading coefficient can be lower bounded as \cite{sheikholeslami2015jamming}:
\small
\begin{align}\nonumber
C_{S,D}\hspace{-2pt}&\left(|h_{S,D}|^2\right)\hspace{-2pt}
\geq\hspace{-2pt}\log\hspace{-2pt}\left(\hspace{-2pt}\frac{\frac{P_{S}|h_{S,D}|^2}{d_{S,D}^{\alpha}}+\sigma_J^2+\sigma_D^2+\frac{\delta^{2}_D}{12}}{\sigma_J^2+\sigma_D^2+\frac{\delta^{2}_D}{12}}\hspace{-2pt}\right)\hspace{-2pt},\\ &=\hspace{-2pt}\log\hspace{-2pt}\left(\hspace{-2pt}\frac{\frac{P_{S}|h_{S,D}|^2}{d_{S,D}^{\alpha}}+\frac{\theta^2P_{J}|h_{S,D}|^2}{d_{S,D}^{\alpha}}+\sigma_D^2+\frac{\delta^{2}_D}{12}}{\frac{\theta^2P_{J}|h_{S,D}|^2}{d_{S,D}^{\alpha}}+\sigma_D^2+\frac{\delta^{2}_D}{12}}\hspace{-2pt}\right)\hspace{-2pt},
\end{align}
\normalsize
and the capacity of the channel between the transmitter $S$ and the eavesdropper $E$ can be upper bounded as \cite{sheikholeslami2015jamming}:
\begin{equation}
C_{S,E}\left( |h_{S,E}|^2\right) \leq\log\left(\frac{\frac{P_{S}|h_{S,E}|^2}{d_{S,E}^{\alpha}}+\sigma_E^2+\frac{\delta^{'2}_E}{12}}{\sigma_E^2+\frac{\delta^{'2}_E}{2\pi e}}\right).
\end{equation}


In order to guarantee proper signal reception at the legitimate receiver, the capacity of the main channel should be greater than a predetermined threshold $\gamma_D^*$.
Let us define,
\begin{equation}\label{eq:gd}
\gamma_D=\frac{\frac{P_{S}|h_{S,D}|^2}{d_{S,D}^{\alpha}}+\frac{\theta^2P_{J}|h_{S,D}|^2}{d_{S,D}^{\alpha}}+\sigma_D^2+\frac{\delta^{2}_D}{12}}{\frac{\theta^2P_{J}|h_{S,D}|^2}{d_{S,D}^{\alpha}}+\sigma_D^2+\frac{\delta^{2}_D}{12}}.
\end{equation}
Hence, the communication between source and destination is reliable if,
\begin{align}\label{eq:gamma_d}
\gamma_D\geq\gamma_D^*.
\end{align}
We define the average  outage probability between $S$ and $D$ as,
\begin{equation}
p_{out}=\mathbb{P}\left( \gamma_D<\gamma_D^*\right).
\end{equation}
In order to guarantee secrecy, the capacity of the channel between the transmitter and eavesdropper should be less than a predetermined threshold $\gamma_E^*$. We define,
\begin{align}
\gamma_E=\frac{\frac{P_{S}|h_{S,E}|^2}{d_{S,E}^{\alpha}}+\sigma_E^2+\frac{\delta^{'2}_E}{12}}{\sigma_E^2+\frac{\delta^{'2}_E}{2\pi e}}.
\end{align}
Hence, the communication between source and destination is secure if,
\begin{align}\label{eq:gamma_e}
\gamma_E<\gamma_E^*.
\end{align}
We define the average secrecy-outage probability (i.e. eavesdropping probability) as,
\begin{equation}
p_{eav}=\mathbb{P}\left( \gamma_E\geq \gamma_E^*\right).
\end{equation}

From (\ref{eq:gamma_d}) and (\ref{eq:gamma_e}) we conclude that if reliability and secrecy constraints are satisfied, the secrecy rate of at least,
 \begin{align}\label{eq:rs}
 R_s=\log(\gamma_D^*)-\log(\gamma_E^*),
 \end{align} 
 can be achieved.
However  as described above, instead of considering a constraint on the secrecy rate,  we consider constraints on the  individual success probabilities of the  receiver and the eavesdropper.
If we instead put the constraint on the secrecy rate,  for a single secrecy rate many  $(\gamma_D,\gamma_E)$ would satisfy the constraint. 
But codes are designed to work on a specific $(\gamma_D, \gamma_E)$ \cite{dehghan2012energy}; hence, we consider (\ref{eq:gamma_d}) and (\ref{eq:gamma_e}) as our reliability and secrecy constraints, respectively.


\section{SERJ: Secure Energy-efficient Routing using Jamming}\label{sec:analysis}
Consider multi-hop communication between two arbitrary nodes, source $S$ and destination $D$.
Suppose $\bold{\Pi}_{SD}$ denotes the set of \textit{all possible paths} between source  $S$ and destination   $D$, and $\Lambda(.)$ is the  cost of communication. 
Our goal is to find the optimum path $\Pi^*$ from the set $\bold{\Pi}_{SD}$ such that,
\[
\Pi^*=\arg\min_{\Pi\in\bold{\Pi}_{SD}}\Lambda\left(\Pi\right),
\]
Please note that for a path $\Pi$, $\Lambda(\Pi)$ is the total cost of secret communication, which  consists of the power
to transmit the message $P_{S_i}$ and the jamming power $P_{J_i}$ of each transmitter along the path $\Pi$, i.e. our optimization objective is,
\begin{align}\label{eq:opt}
\Lambda\left(\Pi\right)=\min \sum_{\ell_i\in\Pi}(P_{S_i}+P_{J_i}).
\end{align}
where the optimization is over all paths in $\bold{\Pi}_{SD}$ and all $P_{S_i}$s and $P_{J_i}$s of the transmitters along the optimum path. By applying the coding technique described in \cite{koyluoglu2012secrecy}, securing each hop is sufficient to ensure end-to-end
secrecy. 
Hence, we consider the following secrecy constraints,
\begin{align}\label{eq:sec_cons}
	 \gamma_{E_{i,j}}<\gamma_{E}^*, \; \forall {\ell_i\in\Pi}\;\text{and}\;\forall E_j\in\mc{E},
\end{align}
which means that for all  eavesdroppers $E_j\in\mc{E}$ in the network, and for all links ${\ell_i}$ along the path $\Pi$, the secrecy constraint must be satisfied. In other words, the communication of each  link ${\ell_i\in\Pi}$ must be secure from every and all eavesdroppers in the network.

Transmission is reliable provided that the following end-to-end average outage probability constraint is guaranteed, 
\begin{align}\label{eq:cons_rel}
	p_{OUT}^{SD}=1-\prod_{\ell_i\in\Pi}{\left(1-p_{out}^{i}\right)}\leq \epsilon.
\end{align}
where $p_{out}^{i}$ denotes the average outage probability of the link $\ell_i=\left\langle S_i,D_i\right\rangle$. Also, the following constraints should be satisfied,
\begin{align}
	\label{eq:cons} P_{S_i}\geq 0,\; \text{and} \;P_{J_i}\geq 0.
\end{align}


\subsection{Analysis of Secrecy}

Consider the secrecy constraint (\ref{eq:sec_cons}).
Substituting $\delta'_E$ from (\ref{eq:delta'}) into (\ref{eq:gamma_e}), $\gamma_{E_{i,j}}$ can be written as,
\begin{align}
\gamma_{E_{i,j}}
&=\frac{\frac{P_{S_i}|h_{S_i,E_j}|^2}{d_{S_i,E_j}^{\alpha}}\left(1+\frac{4l^2}{12\times2^{2b_E-2K_i}}\right)+\sigma_{E}^2}{\frac{P_{S_i}4l^2|h_{S_i,E_j}|^2}{2\pi ed_{S_i,E_j}^{\alpha}2^{2b_E-2K_i}}+\sigma_{E}^2}, \label{eq:secrecy1}
\end{align}
 where without loss of generality, we assume that all eavesdroppers use $b_E$-bit A/Ds (or $b_E$ is the highest resolution that the  A/D of an eavesdropper can have). Since we do not want to make assumptions on the eavesdroppers's noise characteristics, we assume $\sigma_{E}^2 = 0$.
 Note that the assumption of $\sigma_{E}^2=0$ is in favor of the eavesdropper, i.e. the secrecy capacity of the wiretap channel with any $\sigma_{E}^2>0$ is  more than the secrecy capacity of the same wiretap channel  with $\sigma_{E}^2=0$. Consequently, if our algorithm is able to provide secrecy when $\sigma_{E}^2=0$, it can also provide secrecy when $\sigma_{E}^2>0$. 
Substituting $\sigma_{E}^2=0$ in (\ref{eq:secrecy1}), the worst case $\gamma_{E_{i,j}}$ can be written as,
\begin{align}
\gamma_{E_{i,j}}\nonumber
&=\frac{\frac{P_{S_i}|h_{S_i,E_j}|^2}{d_{S_i,E}^{\alpha}}\left(1+\frac{4l^2}{12\times2^{2b_E-2K_i}}\right)}{\frac{P_{S_i}4l^2|h_{S_i,E_j}|^2}{2\pi ed_{S_i,E_j}^{\alpha}2^{2b_E-2K_i}}}\\
&=\frac{1+\frac{4l^2}{12\times2^{2b_E-2K_i}}}{\frac{4l^2}{2\pi e2^{2b_E-2K_i}}}<\gamma_{E}^*.\label{eq:secrecy2}
\end{align}
In (\ref{eq:secrecy2}) $\gamma_{E_{i,j}}$ does not depend on the eavesdroppers's noise $\sigma_{E}^2$ (since we have considered the worst case), the eavesdroppers's location and the eavesdroppers's channel state information (CSI). Thus, if we ensure $\gamma_{E_{i,j}}<\gamma_{E}^*$, the transmission of the relay $S_i$ will be secure from the eavesdropper $E_j$ regardless of noise, location and CSI of $E_j$. 
Further,  note that $\gamma_{E_{i,j}}$ in (\ref{eq:secrecy2}) is a deterministic function, and does not depend on the probabilistic nature of the channel (it does not depend on  $|h_{S_i,E_j}|^2$), and thus if (\ref{eq:secrecy2}) is satisfied, the probability that an arbitrary eavesdropper in the network intercepts the message transmitted by $S_i$ is zero.
This means that if we choose $K_i$ such that (\ref{eq:secrecy2}) is satisfied, none of the eavesdroppers in the network $E_j\in\mc{E}$ can intercept the message that $S_i$ is transmitting.
Rearranging (\ref{eq:secrecy2}), the number of key bits per jamming symbol $K_i$ is lower bounded as,
\begin{align}\label{eq:key}
	K_i>\frac{1}{2}\log_2\left(\frac{\pi e 2^{2b_E-1}}{l^2(\gamma^*_E-\pi e/6)}\right).
\end{align}
 This bound only depends on the resolution of the eavesdropper's
 A/D (which is assumed to be bounded by $b_E$,
 as discussed in Section \ref{sec:approach}), and does not depend on the
 eavesdropper's location or its CSI. Intuitively, when the
 number of key bits per jamming symbol is sufficiently
 large, the quantization noise becomes large enough to
 protect the message against the eavesdropper regardless
 of its location or its CSI.
 Since the lower bound of $K_i$ in (\ref{eq:key}) does not depend on the characteristics of a specific transmitter $S_i$, we can infer that  the same number
 of key bits per jamming symbol $K$ can be used by all
 transmitters, where the minimum value of  $K$ to guarantee secrecy is,
\begin{align}\label{eq:K}
	K=\left\lceil \frac{1}{2}\log_2\left(\frac{\pi e 2^{2b_E-1}}{l^2(\gamma^*_E-\pi e/6)}\right)\right\rceil.
\end{align}
Note that $\gamma^*_E$ is a parameter that is determined by the wiretap code design and hence we can ensure $\gamma^*_E > \pi e/6$ by adding enough randomness to the codebook.

Now let us consider the optimization objective in (\ref{eq:opt}) again. From (\ref{eq:PJ}), the jamming power at each node is proportional to the transmit power, i.e. $P_{J_i}=\beta_i P_{S_i}$, where,
\begin{align}
\nonumber\beta_i&=\frac{2l^2\left(2^{2K_i+1}-3\times2^{K_i}+1\right)}{3}\\
&=\frac{2l^2\left(2^{2K+1}-3\times2^{K}+1\right)}{3}
\end{align}
Since $\beta_i$ does not depend on a specific transmitter $S_i$, we can have the same $\beta_i$ for all transmitters and write $\beta=\beta_i$. Hence, the relationship between the jamming power at each node and the transmit power is $P_{J_i}=\beta P_{S_i}$, where $\beta$ is a constant.
The secrecy objective in (\ref{eq:opt}) can be written as,
\begin{align}\label{eq:opt2}
\nonumber&\min \sum_{\ell_i\in\Pi}(P_{S_i}+P_{J_i})\\
\nonumber&=\min \sum_{\ell_i\in\Pi}(1+\beta)P_{S_i}\\
&=(1+\beta)\min \sum_{\ell_i\in\Pi}P_{S_i}.
\end{align}
where the last equality follows because $\beta$ is independent of the transmitter, and is already minimized by choosing the minimum $K$ that satisfies the secrecy constraint in (\ref{eq:K}). The optimization in (\ref{eq:opt2}) is  over all possible paths between source $S$ and destination $D$, and over all transmit powers of the transmitters along the optimum path. 
Further, the optimization objective in (\ref{eq:opt2}) can be written as,
\begin{align}\label{eq:optimization}
	\min \sum_{\ell_i\in\Pi}P_{S_i}.
\end{align}
\subsection{Analysis of Reliability}
Now consider  $\gamma_D$ in (\ref{eq:gd}) and the reliability constraint (\ref{eq:cons_rel}).
Without loss of generality, we assume all  legitimate receivers have the same quality A/Ds with zero-mean uniform quantization noise with variance $\frac{\delta_D^2}{2}$, and experience AWGN with the same variances $\sigma_D^2$. For the reliability constraint in (\ref{eq:cons_rel}),
the probability of outage at $D_i$ is,
\small
\begin{align}\nonumber
	&p_{out}^{i}=\mathbb{P}\bigg(\frac{\frac{P_{S_i}|h_{S_i,D_i}|^2}{d_{S_i,D_i}^{\alpha}}+\theta^2\frac{P_{J_i}|h_{S_i,D_i}|^2}{d_{S_i,D_i}^{\alpha}}+\sigma_D^2+\frac{\delta^{2}_D}{12}}{\theta^2\frac{P_{J_i}|h_{S_i,D_i}|^2}{d_{S_i,D_i}^{\alpha}}+\sigma_D^2+\frac{\delta^{2}_D}{12}}\hspace{-2pt}<\hspace{-2pt}\gamma_D^*\hspace{-2pt}\bigg)\\
\nonumber	
&=\mathbb{P}\bigg(\frac{\frac{P_{S_i}|h_{S_i,D_i}|^2}{d_{S_i,D_i}^{\alpha}}+\theta^2\frac{\beta P_{S_i}|h_{S_i,D_i}|^2}{d_{S_i,D_i}^{\alpha}}+\sigma_D^2+\frac{\delta^{2}_D}{12}}{\theta^2\frac{\beta P_{S_i}|h_{S_i,D_i}|^2}{d_{S_i,D_i}^{\alpha}}+\sigma_D^2+\frac{\delta^{2}_D}{12}}\hspace{-2pt}<\hspace{-2pt}\gamma_D^*\hspace{-2pt}\bigg)\\
	&=\mathbb{P}\bigg(|h_{S_i,D_i}|^2<\frac{\left(\gamma^*_D-1\right)\big(\sigma_D^2+\frac{\delta^{2}_D}{12}\big)}
{P_{S_i}(1-(\gamma_D^*-1)\theta^2\beta)/d_{S_i,D_i}^{\alpha}}\bigg)\label{eq:21}\\
	&=1-e^{-\frac{\left(\gamma^*_D-1\right)\left(\sigma_D^2+\frac{\delta^{2}_D}{12}\right)}{P_{S_i}\left(1-\left(\gamma_D^*-1\right)\theta^2\beta\right)/d_{S_i,D_i}^{\alpha}}},\label{eq:17}
\end{align}
\normalsize
where   (\ref{eq:21}) holds given that $1-\left(\gamma_D^*-1\right)\theta^2\beta>0$. Otherwise it is easy to show that $p_{out}^i=1$. Hence, reliable communication is possible if $\theta$ and $\beta$ are small enough. 
The last equality follows because, for Rayleigh fading, $|h_{S_i,D_i}|^2$ is exponentially distributed.
Substituting (\ref{eq:17}) into (\ref{eq:cons_rel}), the end-to-end outage probability constraint is,
\begin{align*}\nonumber
	&p_{OUT}^{SD}
	=1-\prod_{\ell_i\in\Pi}e^{-\frac{\left(\gamma^*_D-1\right)\left(\sigma_D^2+\frac{\delta^{2}_D}{12}\right)}{P_{S_i}\left(1-\left(\gamma_D^*-1\right)\theta^2\beta\right)/d_{S_i,D_i}^{\alpha}}}\\
	& =1-\exp\bigg( {-\sum_{\ell_i\in\Pi}\frac{(\gamma^*_D-1)\big(\sigma_D^2+\frac{\delta^{2}_D}{12}\big)}{P_{S_i}\left(1-(\gamma_D^*-1)\theta^2\beta\right)/d_{S_i,D_i}^{\alpha}}}\bigg)\\
	& \leq \epsilon.
\end{align*}
Thus, the end-to-end reliability constraint turns into,
\begin{align}\label{eq:rel_cons2}
	\sum_{\ell_i\in\Pi}\frac{d_{S_i,D_i}^{\alpha}}{P_{S_i}}\leq\eta,
\end{align}
where, 
\begin{align}\label{eq:eta}
\eta=\frac{\log\big(\frac{1}{1-\epsilon}\big)\left(1-\left(\gamma_D^*-1\right)\theta^2\beta\right)}{(\gamma^*_D-1)\big(\sigma_D^2+\frac{\delta^{2}_D}{12}\big)}.
\end{align} 
\subsection{Optimal Cost of a Given Path}

Our goal is to find the optimal path, which requires the  minimum
transmission and jamming power  to satisfy both outage and reliability constraints.
The optimal path is not known in advance. 
Hence, first we find the optimal transmit and jamming power
allocation for a given path $\Pi$, and then we use it to
design a routing algorithm that finds the optimal path. 
From (\ref{eq:optimization})-(\ref{eq:eta}), in order to find the optimal transmit and jamming power
allocation for a given path, we should solve the following optimization problem,
\begin{align}\label{eq:20}
	\min\limits_{P_{S_i}\geq 0} \sum_{\ell_i\in\Pi}P_{S_i}
\end{align}
subject to:
\begin{align}\label{eq:51}
		\sum_{\ell_i\in\Pi}\frac{d_{S_i,D_i}^{\alpha}}{P_{S_i}}\leq\eta
\end{align}
The left side of (\ref{eq:51}) is a decreasing function of $P_{S_i}$ and our goal is to find the minimum $P_{S_i}$.
Hence, we can substitute the inequality with an equality,
\begin{align}\label{eq:1}
\sum_{\ell_i\in\Pi}\frac{d_{S_i,D_i}^{\alpha}}{P_{S_i}}=\eta
\end{align}
This optimization problem can be solved using the technique of Lagrange multipliers.
We must solve  (\ref{eq:20}) and the following equations simultaneously, 
\begin{align*}
	&\frac{\partial}{\partial{P_{S_i}}}\left\{\sum_{\ell_i\in\Pi}P_{S_i}+\lambda\left(\sum_{\ell_i\in\Pi}\frac{d_{S_i,D_i}^{\alpha}}{P_{S_i}}-\eta\right)\right\}=0,\\
	&\quad \text{for}\quad i=1,\ldots,H.
\end{align*}
Taking  derivatives we have,
\begin{align}
	1-\lambda{\frac{d_{S_i,D_i}^{\alpha}}{P_{S_i}^2}}=0, \; i=1,\ldots,H,
\end{align}
and thus,
\begin{align}\label{eq:2}
	P_{S_i}=\sqrt{\lambda d_{S_i,D_i}^{\alpha}}
\end{align}
Substituting $P_{S_i},\: i=1,\ldots,H$ from  (\ref{eq:2}) into (\ref{eq:1}), we obtain that,
\begin{align}\label{eq:3}
		\lambda=\frac{1}{\eta^2}\left(\sum_{\ell_k\in\Pi}\sqrt{d_{S_i,D_i}^{\alpha}}\right)^2
\end{align}
Substituting $\lambda$ from    (\ref{eq:3}) into (\ref{eq:2}), the optimal transmit power at each link is given by,
\begin{align}\label{eq:power}
	P_{S_i}=\frac{1}{\eta}\sqrt{ d_{S_i,D_i}^{\alpha}}\sum_{\ell_k\in\Pi}\sqrt{ d_{S_k,D_k}^{\alpha}}
\end{align}
Hence, the aggregate cost of transmitting the message is,
\begin{align}
\sum_{\ell_i\in\Pi}P_{S_i}=\frac{1}{\eta}\left(\sum_{\ell_k\in\Pi}\sqrt{ d_{S_k,D_k}^{\alpha}}\right)^2,
\end{align}
and the cost of jamming is,
\begin{align}
\sum_{\ell_i\in\Pi}P_{J_i}=\frac{\beta}{\eta}\left(\sum_{\ell_k\in\Pi}\sqrt{ d_{S_k,D_k}^{\alpha}}\right)^2.
\end{align}
The  minimum total (signal+jamming) cost of establishing $\Pi$ is,
\begin{align}\label{eq:cost}
	\Lambda\left(\Pi\right)=\frac{1+\beta}{\eta}\left(\sum_{\ell_k\in\Pi}\sqrt{ d_{S_k,D_k}^{\alpha}}\right)^2.
\end{align}
Note that (\ref{eq:cost}) is the minimum cost of establishing an arbitrary path $\Pi\in \bold{\Pi}_{SD}$, where $\bold{\Pi}_{SD}$ is the set of all paths between $S$ and $D$.
\subsection{Routing Algorithm}\label{sec:routing}
Since $\Lambda\left(\Pi\right)$ in (\ref{eq:cost}) is the minimum cost that can be assigned to any path $\Pi\in \bold{\Pi}_{SD}$ between $S$ and $D$, when we want to find the minimum energy path between $S$ and $D$ we should use $\Lambda\left(\Pi\right)$ as our path cost.
Hence, based on the cost $\Lambda\left(\Pi\right)$,  we  assign weights $\mc{W}(\ell_k)$ to each link $\ell_k$ in the network so that the weight of a path $\mc{W}(\Pi)$ is given as the sum of its link weights, i.e.  $\mc{W}(\Pi)=\sum_{\ell_i\in\Pi}\mc{W}(\ell_k)$,
which facilitates constructing a fast routing algorithm, as follows.
These weights should be chosen such that the path $\Pi^*$ that minimizes the  weight of the path $\mc{W}(\Pi^*)$ also minimizes  the  cost of the path $\Lambda\left(\Pi^*\right)$ over all paths in $\bold{\Pi}_{SD}$, i.e. the path with minimum weight is exactly the path with  minimum cost.
Let us define the weight of a link $\ell_k$ between two arbitrary nodes $S_k$ and $D_k$ ($\ell_k=\left\langle S_k,D_k\right\rangle$) in the network as,
\begin{equation}\label{eq:weights}
\mc{W}(\ell_k)=\sqrt{ d_{S_k,D_k}^{\alpha}}.
\end{equation}
Thus, the weight of a path $\Pi$ will be,
\begin{equation}\label{eq:weight_path}
\mc{W}(\Pi)=\sum_{\ell_k\in\Pi}\sqrt{ d_{S_k,D_k}^{\alpha}}.
\end{equation}
Clearly, a path between $S$ and $D$ (in $\bold{\Pi}_{SD}$) that minimizes $\mc{W}(\Pi)$ in (\ref{eq:weight_path}) also minimizes $\Lambda\left(\Pi\right)$ in (\ref{eq:cost}).
Hence, we should assign the weight $\mc{W}(\ell_k)$ described in (\ref{eq:weights}) to any link $\ell_k$ in the network, and  apply any shortest path algorithm like Dijkestra to find the path with minimum weight $\Pi^*$  between $S$ and $D$, which is also the path with minimum cost (the minimum energy path).
%
From (\ref{eq:power}), each node along $\Pi^*$ forwards the message to the next node with total (transmit and jamming) power, 
\begin{align}
\Lambda\left(\ell_i\right)=\frac{1+\beta}{\eta}\sqrt{ d_{S_i,D_i}^{\alpha}}\sum_{\ell_k\in\Pi^*}\sqrt{ d_{S_k,D_k}^{\alpha}}.
\end{align}
 and the total end-to-end cost of communication is,
 \begin{align}
 \Lambda\left(\Pi^*\right)=\frac{1+\beta}{\eta}\left(\sum_{\ell_k\in\Pi^*}\sqrt{ d_{S_k,D_k}^{\alpha}}\right)^2.
 \end{align}
 

\section{Comparison to Previous Work}\label{sec:numerical}

In this section we compare the performance of our algorithm with that of the SMER algorithm \cite{ghaderi2014min} in different scenarios.  

\begin{figure}
	\centering
	\includegraphics[width=.5\textwidth]{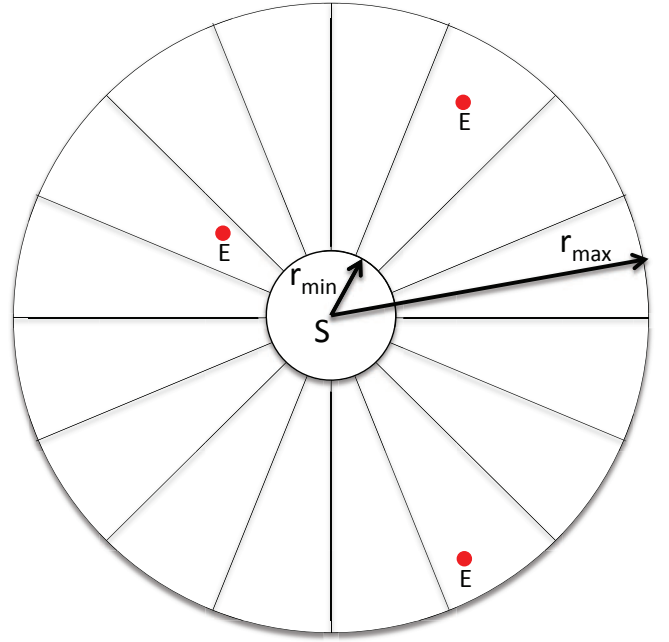}
	\caption{Three sectors have an eavesdropper with probability one, and the rest of the sectors have an eavesdropper with probability zero.}
	\label{fig:smer}
\end{figure}

\textbf{SMER Algorithm.} In SMER, the system nodes employ cooperative jamming to establish a secure path, and, if the eavesdroppers get very close to a transmitter, the secrecy is compromised.  Hence, while the SERJ algorithm proposed here has no need or sense of a ``guard region'', to employ SMER  we must introduce such into the scenario.  
Thus, for the sake of comparison to SMER, assume a guard region with  radius $r_{min}>0$ around each transmitter and assume that no  eavesdropper can enter the guard regions.
Further, in SMER a set of locations and the probability that an eavesdropper exists in each location must be known.
In order to address this requirement of SMER, we divide a circle centered at the transmitter $S$ and  with radius $r_{max}$ into many sectors. 
Each sector is a location where an eavesdropper might exist.
For instance, when three eavesdroppers are present, three sectors have an eavesdropper with probability one, and the rest of the sectors have an eavesdropper with probability zero (Fig. \ref{fig:smer}). 
Unlike  the SERJ algorithm  proposed in this paper, the secrecy outage probability of SMER is non-zero. 
In the next section, we will see how this non-zero eavesdropping probability affects the power consumption of secret communication.

Before we proceed to the numerical results, we compare the asymptotic  complexity of SERJ and SMER algorithms in a network that consists of $n$ system nodes.

\textbf{Running Time.}
In order to find the optimal path using SERJ we should simply assign the weights described in Section \ref{sec:routing} to the links between the legitimate nodes of the network, and then we  need to apply the Dijkstra's algorithm once, which is a polynomial algorithm with running time  $O(n^2)$.
Hence, the asymptotic running time of SERJ is  polynomial in $n$ which makes us classify SERJ as an efficient routing algorithm.
On the other hand, SMER is a pseudo-polynomial algorithm of order $O(n^2B)$, where $B$ is the maximum cost of any path in the network. 
Note that, while the running time of SMER is polynomial in $B$, the 
actual value of $B$ grows exponentially with the size of the input 
(i.e., the number of bits used to represent link costs). That is, if $l$ 
bits are used to represent the link cost values then $B$ will be of 
order $2^l$.
Therefore, in practice, SERJ will be much faster than SMER, especially in situations that the cost of communication is high and thus the value of  $B$ is large (e.g. large networks, large path-loss exponents, high uncertainty in the locations of the eavesdroppers, ...).

To get more insight into the problem, first we consider secure one-hop transmission from a transmitter $S$ to a receiver $D$ in the presence of eavesdroppers.
Next, we will consider multi-hop minimum energy routing in a network and in the presence of multiple eavesdroppers.
In both cases, we assume that the system nodes and the eavesdroppers use 14-bit A/Ds, and we set $\theta=10^{-6}$.
We set the source-destination outage probability $\pi=0.1$, receiver noise power $N_0=1$ (eavesdropper noise power is zero in both SERJ and SMER), $\gamma_D^*=42$ and $\gamma_E^*=34$, which results in the secrecy rate $R_s=0.3$ (bits/use). We consider different propagation attenuation scenarios: $\alpha=2$ which is the path-loss exponent corresponding to  free space, and $\alpha=3$ and $\alpha=4$ which are the path-loss exponents corresponding to a terrestrial environment.

\subsection{One-Hop Communication}

\begin{figure}
	\centering
	\includegraphics[width=.5\textwidth]{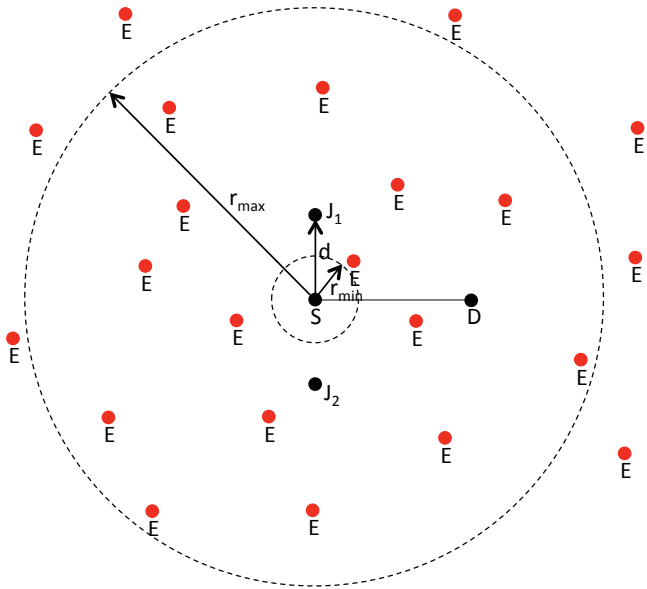}
	\caption{One-hop communication between source $S$ and destination $D$ in the presence of eavesdroppers ($E$s). In SMER, two jammers $J_1$ and $J_2$ help to make the link secure.}
	\label{fig:one_hop2}
\end{figure}

Consider a single hop in a wireless network, consisting of a  transmitter $S$ and a receiver $D$ (Fig. \ref{fig:one_hop2}).
For SMER, suppose  two jammers $J_1$ and $J_2$ help the transmitter  to convey its message to the receiver securely \cite{ghaderi2014min}.
The distance between each jammer and the source is denoted by $d$. 
In the remainder of this section, we consider the effect of various parameters of the network  on the energy consumption of our scheme and SMER.

\begin{figure}
	\centering
	\includegraphics[width=.5\textwidth]{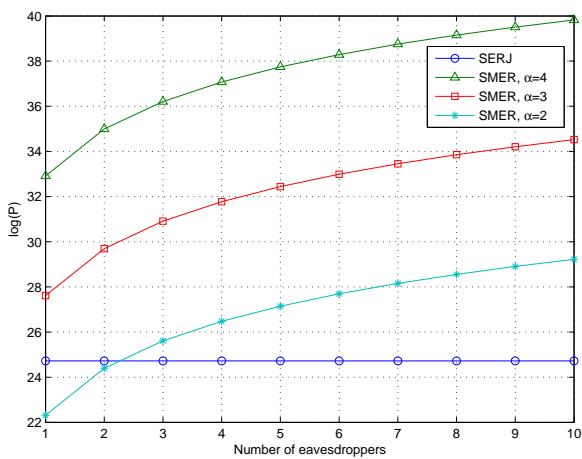}
	\caption{Power consumption of SERJ and SMER versus  the number of eavesdroppers for various values of path-loss exponent $\alpha$. 
		}
	\label{fig:P_vs_numeavesdropper_b14}
\end{figure}
\textbf{Number of Eavesdroppers.}
Fig. \ref{fig:P_vs_numeavesdropper_b14} shows the transmission power versus  the number of eavesdroppers around the transmitter\footnote{In all  figures in this section, $P$ denotes the aggregate power consumed by the algorithm.}. In this figure, $p_{eav}=10^{-5}$, $r_{min}=0.01$, $r_{max}=2$, and $d_{SD}=1$. As   shown in Section \ref{sec:analysis}, the power required when employing  SERJ does not depend on the number of eavesdroppers. On the other hand, when the number of eavesdroppers increases, the power needed to establish a secure link using SMER increases dramatically.
Since the cost of communication using  SERJ only depends on the distance between the transmitter and the receiver which is normalized to  $d_{SD}=1$, the cost of using  SERJ does not change with the change of  path-loss exponent in these plots.

\begin{figure}
	\centering
	\includegraphics[width=.5\textwidth]{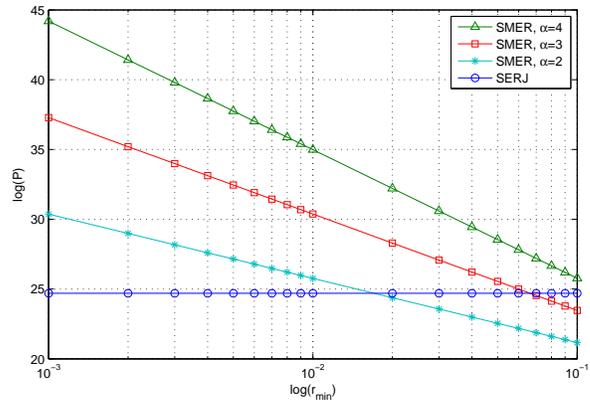}
	\caption{Power consumption of SERJ and SMER versus the radious $r_{min}$ of the guard region  for various values of $\alpha$ and when only one eavesdropper is present. As we allow the eavesdropper to become closer to the transmitter  (i.e. as $r_{min}$ gets smaller), the power needed to make the  link secure using SMER becomes higher. On the other hand, with SERJ there is no need to assume a guard region around the transmitter.}
	\label{fig:P_vs_rmin_b14}
\end{figure}

\textbf{Guard Region Radius.}
Whereas the proposed algorithm (SERJ) does not require a guard region, recall that SMER cannot be utilized without such.
Fig. \ref{fig:P_vs_rmin_b14} shows the   power versus $r_{min}$  in the presence of $n_E=5$ eavesdroppers, and for various values of the path-loss exponent $\alpha$. 
We set   $d_{SD}=1$, $p_{eav}=10^{-5}$ and $r_{max}=2$.  
We observe that when $r_{min}$ gets small, the power needed to establish a secure link using SMER increases dramatically, while the power needed to establish a secure link using SERJ does not depend on the  location of the eavesdropper. 
In fact as  is shown in  Section \ref{sec:analysis}, the power used by SERJ is independent of the distance between the transmitter  and  the eavesdroppers, and, even if the eavesdroppers get very close to the transmitter, they cannot intercept the message.

\begin{figure}
	\centering
	\includegraphics[width=.5\textwidth]{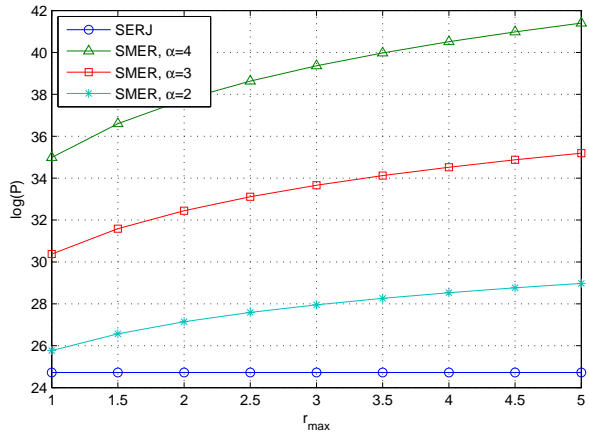}
	\caption{Power consumption of SERJ and SMER versus  $r_{max}$ for various values of $\alpha$ and when  $n_E=5$ eavesdroppers are present. 
		 The performance of SMER  is closely dependent on the uncertainity in the locations of the eavesdroppers, while the performance of SERJ does not depend on the locations of the eavedroppers.}
	\label{fig:P_vs_rmax_b14}
\end{figure}

\textbf{Uncertainty in the Location of  Eavesdroppers.}
In Fig. \ref{fig:P_vs_rmax_b14}, the power needed to transmit the message securely versus $r_{max}$  for various values of the path-loss exponent $\alpha$ is depicted. 
For SMER we set $p_{eav}=10^{-5}$ and $r_{min}=0.01$.
As $r_{max}$ increases, the uncertainty in the location of the eavesdroppers increases, and thus in SMER the jammers need to consume more power to cover a larger area.
On the other hand, with SERJ, the transmit power is independent of the locations of the eavesdroppers.

\begin{figure}
	\centering
		\includegraphics[width=.5\textwidth]{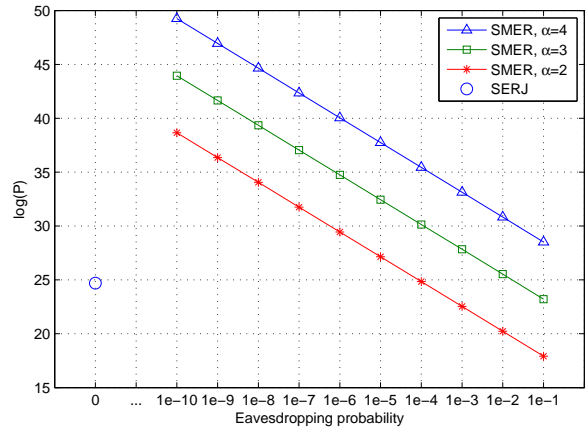}
	\caption{Power consumption of SERJ and SMER versus eavesdropping probability for various values of $\alpha$ and when  $n_E=5$ eavesdroppers are present.   For small secrecy outage probabilities, the power consumption of SMER is substantially higher than the power consumption of  SERJ. }
	\label{fig:P_vs_peav_b14}
\end{figure}

\textbf{Eavesdropping Probability.}
As  was shown in Section \ref{sec:analysis}, the eavesdropping probability of SERJ is zero.
But, the eavesdropping probability of SMER is not zero. 
Fig. \ref{fig:P_vs_peav_b14} shows the power needed to establish a secure link versus the
eavesdropping probability when  $n_E=5$ eavesdroppers are present,  $r_{min}=.01$, and $r_{max}=2$.
It can be seen that the power consumption of SMER dramatically changes when the secrecy outage probability changes. In particular,  for small secrecy outage probabilities, the power consumption of SMER is substantially higher than the power consumption of  SERJ.

\begin{figure}
	\centering
	\includegraphics[width=.5\textwidth]{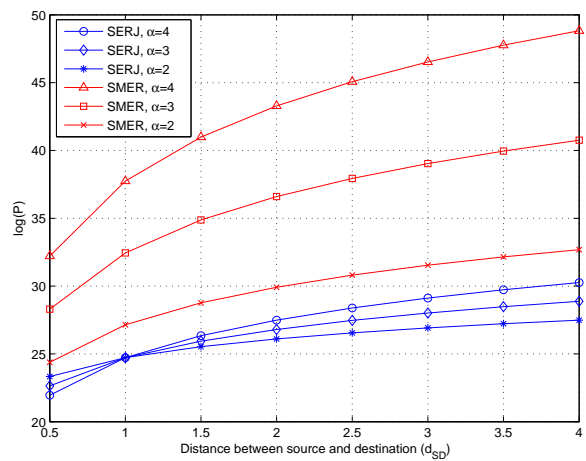}
	\caption{Power consumption of SERJ and SMER versus the distance between source and destination  $d_{SD}$ for various values of $\alpha$ and when  $n_E=5$ eavesdroppers are present.   As the distance between the transmitter and the receiver gets longer, the transmit power of both schemes increases. }
	\label{fig:P_vs_dSD_b14}
\end{figure}

\textbf{Distance between Source and Destination.}
Fig. \ref{fig:P_vs_dSD_b14} shows the transmission power versus  the distance between source and destination  $d_{SD}$ for various values of $\alpha$. For SMER, we set $p_{eav}=10^{-5}$, $r_{min}=.01$, and $r_{max}=2 d_{SD}$. As the distance between the transmitter and the receiver gets longer, the transmit power of both schemes increases.

\begin{figure}
	\centering
	\includegraphics[width=.5\textwidth]{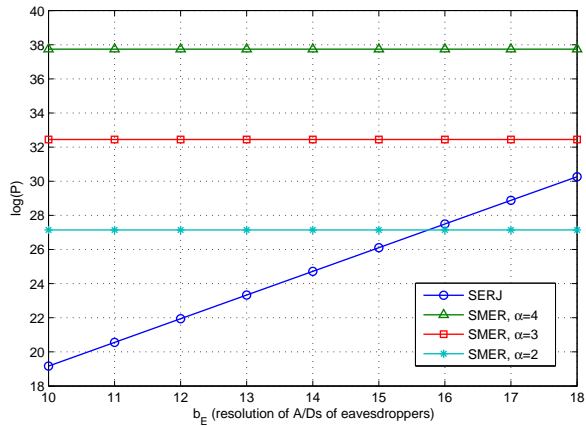}
	\caption{Power consumption of SERJ and SMER versus the resolution of eavesdroppers' A/Ds  $b_E$ for various values of $\alpha$ and when  $n_E=5$ eavesdroppers are present.   As $b_E$ become higher, with SERJ we need more jamming power and thus the power consumption of SERJ increases while with SMER, the performance is independent of the quality of eavesdroppers' A/Ds. }
	\label{fig:P_vs_bE}
\end{figure}

\textbf{Quality of Eavesdroppers' A/Ds.}
Fig. \ref{fig:P_vs_bE} shows the transmission power versus the  resolution of eavesdroppers' A/Ds  $b_E$  for various values of $\alpha$. The distance between the transmitter and the receiver $d_{SD}=1$, and for SMER, we set $p_{eav}=10^{-5}$, $r_{min}=.01$, and $r_{max}=2 $. While with SMER the performance is independent of the quality of eavesdroppers' A/Ds, with SERJ as the  resolution of eavesdroppers' A/Ds gets higher the transmit power  increases because it needs more jamming power to provide secrecy.

\subsection{Multi-Hop Communication}
We consider a wireless network that consists of $n$ system nodes and $n_E$ eavesdroppers which are distributed uniformly at random 
on a  $5\times 5$ square. 
Our goal is to find a secure path with minimum aggregate energy from the source to the destination, using SERJ and SMER.
For the remainder of this section, we assume that in SMER, for every node, two friendly jammers exist that help the node to establish a secure link. 
We average the results over 10 random realizations of the network. In each realization, the system nodes are distributed uniformly at random, and  the closest system node to the point $(0, 0)$
is the source of the message  and the closest system node to the point $(5,5)$
is the destination.
We  consider the path-loss exponent $\alpha=3$, since $\alpha=2$ corresponds to non-terrestrial environments, and  $\alpha=4$ leads to very high link costs of  SMER, which makes the running time of SMER excessively high.
In the sequel, we investigate the effect of various parameters on the total energy consumption of  SERJ and SMER, and compare their performances.

\begin{figure}
	\centering
	\includegraphics[width=.5\textwidth]{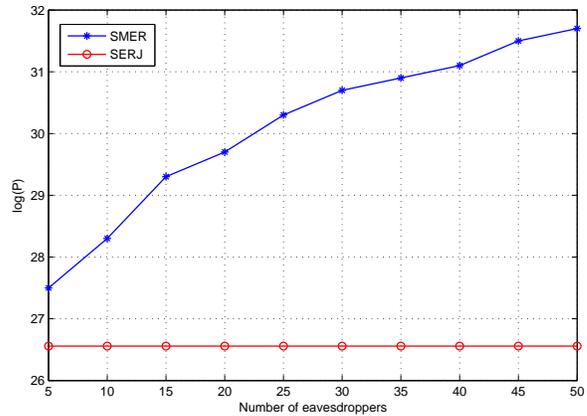}
	\caption{Power consumption of SERJ and SMER versus the number of  eavesdroppers.
		 As the number of eavesdroppers  increases, the amount of power that SMER uses increases, while the amount of power that SERJ uses does not depend on the number and location of the eavesdroppers. }
	\label{fig:network_PVseavesdropper_a3}
\end{figure}

\textbf{Number of Eavesdroppers.}
The average power $P$ versus the number of  eavesdroppers for SERJ and SMER is shown in Fig. \ref{fig:network_PVseavesdropper_a3}.
 There are $n = 25$ system nodes in addition to the eavesdroppers.
 The path-loss exponent of the environment is $\alpha=3$.
For SMER, we set $p_{eav}=10^{-5}$, $r_{min}=.03$, and $r_{max}=2$. 
It can be seen that for very small numbers of eavesdroppers, the performance of SMER is better than that of  SERJ. However, as the number of eavesdroppers  increases, the amount of power that SMER uses increases and becomes more than the  power that SERJ consumes. 
As  is shown in Section \ref{sec:analysis},  the amount of power that SERJ uses does not depend on the number and location of the eavesdroppers.

\begin{figure}
	\centering
	\includegraphics[width=.5\textwidth]{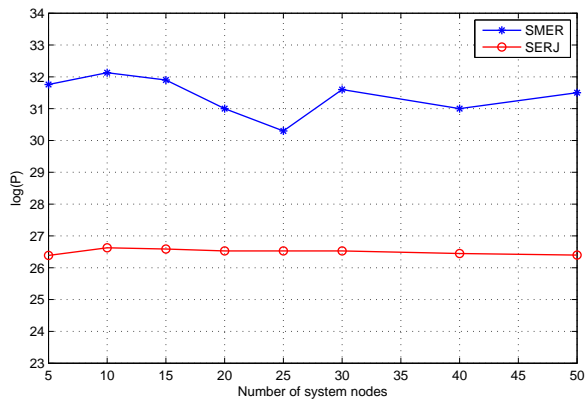}
	\caption{Power consumption of SERJ and SMER versus the number of  system nodes.
		For both algorithms
		the average power is not sensitive to the number of
		system nodes. }
	\label{fig:p_vs_n_a3_rmax2}
\end{figure}

\textbf{Number of System Nodes.}
The effect of the number of system nodes on the average aggregate power consumption   is shown in Fig. \ref{fig:p_vs_n_a3_rmax2}.
There are 
$n_E=25$ eavesdroppers, and the path-loss exponent of the environment is $\alpha=3$.
For SMER, we set  $p_{eav}=10^{-5}$, $r_{min}=.03$, and $r_{max}=2$.
 
It can be seen that the performance of SERJ
 is always superior to the performance of SMER.
  For both algorithms
the average power is not sensitive to the number of
system nodes. 
The fluctuations in this figure are due to the random
generation of  network configurations.

\begin{figure}
	\centering
	\includegraphics[width=.5\textwidth]{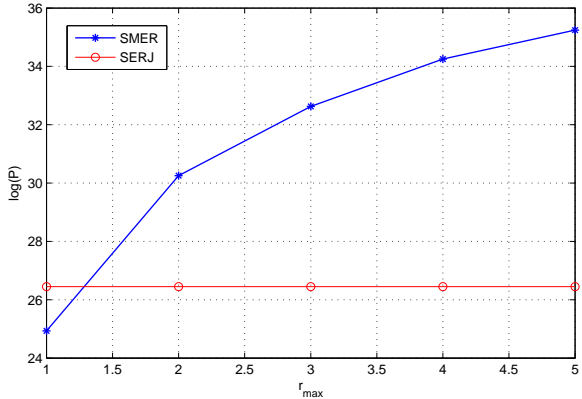}
	\caption{Power consumption of SERJ and SMER versus the uncertainity in the location of the eavesdropper (i.e. $r_{max}$ around each transmitter in the network).
		The transmit power using SERJ is independent of the location of the eavesdroppers.
		But with SMER, as the uncertainty in the location of the eavesdroppers increases the power consumption increases.
		}
	\label{fig:p_vs_rmax_a3_nen25}
\end{figure}

\textbf{Uncertainty in the Location of the Eavesdroppers.}
In Fig. \ref{fig:p_vs_rmax_a3_nen25}, the power needed to transmit the message securely versus $r_{max}$   is shown. 
There are $n=25$ system nodes and 
$n_E=25$ eavesdroppers, and the path-loss exponent of the environment is $\alpha=3$.
For SMER, we set $p_{eav}=10^{-5}$ and $r_{min}=0.03$.
With SERJ, the transmit power is independent of the location of the eavesdroppers.
With SMER, as $r_{max}$ increases, the uncertainty in the location of the eavesdroppers increases, and thus  the jammers need to consume more power to cover a larger area.
For the case that SMER is secure against any eavesdropper in the network (i.e. $r_{max}=5$, if we do not consider the guard regions around the transmitters), the power spent by SMER is substantially higher than the power spent by SERJ. 

\section{Conclusions}\label{sec:conclusions}

In this paper, we have considered secure energy-efficient routing in a quasi-static multi-path fading environment in the presence of passive eavesdroppers. 
Since the eavesdroppers are passive, their locations and CSIs are not known to the legitimate nodes.
Thus we   looked for approaches that do not  rely on the locations and quality of the channels of the eavesdroppers.
We  developed an energy-efficient routing algorithm  based on  random jamming to exploit non-idealities of the eavesdropper's receiver  to provide secrecy. 
Our routing algorithm is fast (finds the optimal path in polynomial time), and does not depend on the number of eavesdroppers and their location and/or channel state information.

We have performed several simulations over single-hop and multi-hop networks with  various network parameters, and compared the performance of our proposed algorithm with that of the SMER algorithm of \cite{ghaderi2013efficient,ghaderi2014min}. 
A major weakness of SMER is that it requires the definition of a guard region that restricts how close eavesdroppers can come 
to system nodes.
Even with such a guard region, which SERJ does not require, we observed that when the uncertainty in the location of the eavesdroppers is high and in disadvantaged wireless environments, the energy consumption of our algorithm  is substantially less than that of the SMER algorithm. 
Gains of SERJ over SMER would be even more substantial in environments with "smart" eavesdroppers; for example, 
eavesdroppers that located themselves close to system nodes or pointed directional antennas at system nodes would significantly
degrade the performance of SMER, but there would be no impact on the performance of SERJ.
Hence, the proposed algorithm directly addresses one of the key roadblocks to the implementation of information-theoretic security in wireless networks: robustness to the operating environment.

\bibliographystyle{ieeetran}
\bibliography{mycite}

\end{document}

%% file: abstract.tex
This paper considers secure energy-efficient routing  in the presence of multiple passive eavesdroppers. Previous work in this area has considered secure routing assuming probabilistic or exact knowledge  of the location and channel-state-information (CSI) of each eavesdropper. In wireless networks, however, the locations and CSIs of passive eavesdroppers are not known, making it challenging to guarantee  secrecy  for any routing algorithm. 

We develop an efficient (in terms of energy consumption and computational complexity) routing algorithm that does not rely on any information about the locations and CSIs of the eavesdroppers. 
Our algorithm  guarantees secrecy even in disadvantaged wireless environments, where multiple eavesdroppers try to eavesdrop  each message, are equipped with directional antennas, or can get arbitrarily close to the transmitter. 
The key is to employ additive random jamming to exploit inherent non-idealities of the eavesdropper's
receiver, which makes the eavesdroppers incapable of  recording the messages.
We have simulated our proposed algorithm and compared it with existing secrecy routing algorithms in both single-hop
and multi-hop networks. Our results indicate that when the uncertainty in the locations of eavesdroppers is high and/or in disadvantaged wireless environments, our algorithm outperforms existing algorithms in terms of energy consumption and secrecy.

%% file: introduction.tex
Information secrecy has traditionally been achieved by cryptography, which is based
on assumptions on current and future computational capabilities of the adversary.
However, there are numerous examples of cryptographic schemes being broken that
were supposedly secure \cite{paar2010understanding}. 
This motivates the consideration of physical layer schemes which are based on information-theoretic secrecy \cite{wyner1975wire}.  
 In a scenario where an adversary tries to eavesdrop on the main channel between a transmitter and a receiver, Wyner showed that,  if the eavesdropper's channel is
degraded with respect to the main channel,  a positive secrecy rate can be achieved.
This idea was later extended to  Gaussian channels \cite{leung1978gaussian}, and to the more general case of a
wiretap channel with a "more noisy" or "less capable" eavesdropper channel \cite{csiszar1978broadcast}.
Thus, the key to obtain information-theoretic secrecy is having an advantage for the main channel against the eavesdropper channel.
However, such an advantage  cannot always be guaranteed.
In particular, the locations of eavesdroppers are not known and an eavesdropper might be much closer
to the transmitter than the intended receiver.
To overcome this problem,  one must design algorithms
to obtain the required advantage for the intended recipient
over the eavesdroppers. 

The idea of
adding artificial noise to the signal by means of  multiple
antennas at the transmitter or  some
helper nodes was introduced in \cite{negi2005secret}. 
The artificial noise is placed in the null
space of the channel from the transmitter to the intended recipient
and thus does not affect it.
But, it  degrades the eavesdropper's channel
with high probability. Subsequently, cooperative jamming 
for physical layer secrecy  has been
extensively studied, e.g.  \cite{tekin2008general,dong2009cooperative,krikidis2010jamming,zheng2011optimal,xie2014secure,gabry2015energy}.
These works mainly focus on one-hop networks consisting of  one transmitter,
one receiver, one eavesdropper and maybe a few helper nodes that generate the artificial noise.
The case of two-hop networks  consisting of  one transmitter,
one receiver, one relay, one eavesdropper and  a few noise generating helper nodes  has also been considered extensively in the literature \cite{lai2008relay,Dennis2011JSAC,kim2012combined,zou2013optimal}. 
In the case of multi-hop networks with multiple transmitters and receivers and in the presence of many eavesdroppers, often the asymptotic results for large networks have been investigated \cite{pinto2008physical,liang2009secrecy,koyluoglu2010secrecy,sheikholeslami2012physical,mirmohseni2014scaling}.

However, whereas one-hop, two-hop and asymptotically large networks are most amenable to analysis and do provide insight into wireless network operation, most ad hoc networks in practice operate with a number of nodes and a number of hops that is between these two extremes.  Hence, the design of algorithms to provide secrecy  in networks of arbitrary "moderate" size is of interest, which is considered here.  
We consider a network with multiple system nodes where a source node communicates with a destination node in a multi-hop fashion and in the presence of multiple passive eavesdroppers. 
We define the cost of communication to be the total energy spent by the system
nodes to securely and reliably transmit a message from the source to the
destination.
Thus, our goal is to find routes that minimize the cost of transmission between the source and destination nodes.
Energy efficiency is an important consideration in designing the routing algorithms, and energy efficient routing has been extensively studied in the literature, e.g. \cite{singh1998power,rodoplu1999minimum,chang2000energy,shah2002energy,kwon2006energy,pandana2008robust,tekbiyik2011energy,zhu2011model,gurakan2015optimal,sheikholeslami2014jamming,sheikholeslami2016energy}.
However, only a few works have considered  energy-aware routing
with secrecy considerations \cite{ghaderi2014min,ghaderi2013efficient}.

In  \cite{ghaderi2014min,ghaderi2013efficient}, the authors use a general probabilistic model for the location of each eavesdropper, and introduce a routing algorithm called SMER (secure minimum energy routing) which employs cooperative jamming to provide  secrecy at each hop such that the  end-to-end secrecy of the multi-hop source-destination path is guaranteed.
When the density of eavesdroppers is low such that there is only one eavesdropper per hop,    the location of the eavesdroppers are known, and the eavesdroppers are restricted to use omni-directional antennas, this approach is promising.
However, since we are considering passive eavesdroppers, their location and channel-state-information (CSI) are not known to the legitimate nodes.
Further, in a  disadvantaged wireless environment,
  many passive eavesdroppers might try to intercept the message at each hop, with large 
uncertainty in the locations of the eavesdroppers, and the eavesdroppers  might get arbitrarily close to the transmitters.
In such a situation, the energy consumption of any cooperative jamming approach including the scheme of  \cite{ghaderi2014min,ghaderi2013efficient} can become  very high.
Further, if we plan for the wrong number of eavesdroppers or do not correctly anticipate the quality of the  eavesdroppers' channels, the secrecy will be compromised.
 Hence, in this paper we seek methods that do not rely on the quality of eavesdroppers' channels and their locations and can provide secrecy in disadvantaged environments at a reasonable cost.
 
 Recently, in \cite{sheikholeslami2012exploiting,sheikholeslami2013everlasting,sheikholeslami2013artificial,sheikholeslami2014everlasting,sheikholeslami2015jamming}, authors have introduced the idea of employing an ephemeral key to exploit imperfections of the eavesdropper's A/D to obtain everlasting secrecy. In contrast to other
 methods based on a key to facilitate secrecy in wireless networks, the works in \cite{sheikholeslami2012exploiting,sheikholeslami2013everlasting,sheikholeslami2013artificial,sheikholeslami2014everlasting,sheikholeslami2015jamming} do not presume that the key is kept secret from the eavesdropper indefinitely; rather, a distortion is used to build an advantage for the intended receiver over the eavesdropper by inhibiting the eavesdropper's ability to even record a reasonable version of the message for later decoding. In particular,
 \cite{sheikholeslami2015jamming} introduced the idea of adding a random jamming signal with large variations based on the ephemeral key to obtain secrecy in disadvantaged wireless environments.   The work of \cite{sheikholeslami2015jamming} considered
 a basic point-to-point communication setting in the presence of one eavesdropper, and thus did not consider the probabilistic behavior of a real communication channel or the impact of imperfections in the channel estimation and jamming cancellation at the intended receiver.   In this paper, we address the application of \cite{sheikholeslami2015jamming} in a multi-hop network  in the presence of multiple eavesdroppers with unknown locations and CSIs.   Also, we consider a more realistic wireless setting than \cite{sheikholeslami2015jamming}, and design an efficient (polynomial time) routing algorithm such that the aggregate energy spent to convey the message and to generate the random jamming signal is minimized.  Hence, a summary of the contributions of this paper is: 
\begin{itemize}
	\item 
	In the modeling of the point-to-point links in our network, we consider a more realistic wireless communication environment compared to the line-of-sight communication considered in the point-to-point method described in \cite{sheikholeslami2015jamming} by:  (a) incorporating multi-path fading in our modeling and analysis; (b) in contrast to secrecy approaches that consider perfect jamming cancellation at the legitimate receiver (e.g. \cite{negi2005secret,sheikholeslami2015jamming}), considering the channel estimation error which causes an error in the cancellation of the jamming signal at the intended receiver.
	\item
We develop an optimization framework to minimize the amount of energy that is used by the random jamming technique to convey a message reliably and securely from a source node to a destination node in a multi-hop fashion. Based on this optimization framework, we provide an efficient routing algorithm that can be used to establish a secure minimum energy path between any pair of nodes in a wireless network with arbitrary node placement.
	\item
	We show that secure and reliable multi-hop communication is possible in an arbitrary network, even in the presence of multiple eavesdroppers of unknown number, locations and CSIs.  Notably, we show that the near eavesdropper challenge, which is a critical challenge in providing physical layer secrecy in wireless networks (e.g. see \cite{ghaderi2014min}), especially in the case of passive eavesdroppers with unknown locations and CSIs, can be resolved using the random jamming technique.
	\item
	We show that the algorithm developed from the random jamming approach coupled with our approach to network optimization:  (a) has improved performance in different scenarios compared to other approaches (i.e. SMER \cite{ghaderi2014min}); (b) has performance that is independent of the particular statistical distribution of the channel gain between the transmitter and the eavesdropper, and thus will work for any kind of eavesdropper's channel.
\end{itemize}

The rest of the paper is organized as follows. Section
\ref{sec:system_and_approach} describes the system model, the approach which is used in this paper, and the  metric. The analysis of the problem and the algorithm for minimum
energy routing with secrecy constraints is presented in
Section \ref{sec:analysis}. In Section \ref{sec:numerical}, the results of numerical
examples for various realizations of one-hop and multi-hop  systems are provided,
and the comparison of the proposed method to SMER algorithm
 is presented. Conclusions  are discussed in Section \ref{sec:conclusions}.